\documentclass{elsart}
\usepackage{amsmath}
\usepackage{color}
\usepackage{amssymb}
\usepackage{graphicx}
\usepackage{subfigure,hyperref}

\newcommand \be{\begin{equation}}
\newcommand \ba{\begin{eqnarray}}
\newcommand \ee{\end{equation}}
\newcommand \ea{\end{eqnarray}}

\begin{document}
\begin{frontmatter}
\title{Statistical properties of daily ensemble variables in the Chinese stock markets}
\author[SB,ICCT]{\small{Gao-Feng Gu}},
\author[SB,SS]{\small{Wei-Xing Zhou}\thanksref{Addr}}
\address[SB]{School of Business,\\East China University of Science and Technology, Shanghai 200237, China}
\address[SS]{School of Science,\\East China University of Science and Technology, Shanghai 200237, China}
\address[ICCT]{Institute of Clean Coal Technology,\\East China University of Science and Technology, Shanghai 200237, China}
\thanks[Addr]{Corresponding author. {\it E-mail address:}\/
wxzhou@moho.ess.ucla.edu}

\begin{abstract}
We study dynamical behavior of the Chinese stock markets by
investigating the statistical properties of daily ensemble returns
and varieties defined respectively as the mean and the standard
deviation of the ensemble daily price returns of a portfolio of
stocks traded in China's stock markets on a given day. The
distribution of the daily ensemble returns has an exponential form
in the center and power-law tails, while the variety distribution is
log-Gaussian in the bulk followed by a power-law tail for large
varieties. Based on detrended fluctuation analysis, R/S analysis and
modified R/S analysis, we find evidence of long memory in the
ensemble returns and strong evidence of long memory in the evolution
of variety.
\end{abstract}

\begin{keyword}
Econophysics, Ensemble return, Variety, Probability distribution,
Long memory, Statistical test
\end{keyword}

\end{frontmatter}

\typeout{SET RUN AUTHOR to \@runauthor}

\section{Introduction}

Financial markets are complex systems, in which participants
interact with each other and react to external news attempting to
gain extra earnings by beating the markets. In the last decades,
econophysics has become to flourish since the pioneering work of
Mantegna and Stanley in 1995 \cite{Mantegna-Stanley-1995-Nature}.
Econophysics is an emerging interdisciplinary field, where theories,
concepts, and tools borrowed from statistical mechanics, nonlinear
sciences, mathematics, and complexity sciences are applied to
understand the self-organized complex behaviors of financial markets
\cite{Mantegna-Stanley-2000,Bouchaud-Potters-2000,Sornette-2003}.
Econophysicists have discovered or rediscovered numerous stylized
facts of financial markets
\cite{Mantegna-Stanley-2000,Cont-2001-QF}, such as fat tails of
return distributions
\cite{Mandelbrot-1963-JB,Mantegna-Stanley-1995-Nature,Ghashghaie-Breymann-Peinke-Talkner-Dodge-1996-Nature,Johansen-Sornette-1998-EPJB,Laherrere-Sornette-1998-EPJB,Gopikrishnan-Meyer-Amaral-Stanley-1998-EPJB,Gopikrishnan-Plerou-Amaral-Meyer-Stanley-1999-PRE,Plerou-Gopikrishnan-Amaral-Meyer-Stanley-1999-PRE,Malevergne-Pisarenko-Sornette-2005-QF},
absence of autocorrelations of returns \cite{Mantegna-Stanley-2000},
long memory in volatility
\cite{Liu-Cizeau-Meyer-Peng-Stanley-1997-PA,Arneodo-Muzy-Sornette-1998-EPJB,Liu-Gopikrishnan-Cizeau-Meyer-Peng-Stanley-1999-PRE},
intermittency and multifractality
\cite{Ghashghaie-Breymann-Peinke-Talkner-Dodge-1996-Nature,Mantegna-Stanley-1996-Nature,Ivanova-Ausloos-1999-EPJB,Mandelbrot-2001b-QF},
and leverage effect
\cite{Bouchaud-Matacz-Potters-2001-PRL,Bouchaud-Potters-2001-PA}, to
list a few.

Recently, Lillo and Mantegna have introduced the conception of
ensemble variable treating a portfolio of stocks as a whole
\cite{Lillo-Mantegna-2000-EPJB,Lillo-Mantegna-2000-PRE,Lillo-Mantegna-2001-EPJB,Lillo-Mantegna-2001-PA}.
They have defined two quantities, the ensemble return and the
variety. The ensemble return ${\mu}$ is the mean of the returns of
the portfolio at time $t$, which is a measure of the market
direction, while the variety ${\sigma}$ is the standard deviation of
all the the returns at time $t$, which characterizes how different
the behavior of stocks is. In the time periods when the markets are
very volatile, the ensemble returns have larger fluctuations and the
varieties are larger as well. It is very interesting to note that
there are sharp peaks in the variety time series when the market
crashes \cite{Lillo-Mantegna-2001-EPJB,Lillo-Mantegna-2001-PA},
which is reminiscent of the behavior of volatility. In addition, the
daily ensemble return of stocks in the New York Stock Exchange is
found to be uncorrelated, while the daily variety has long memory
\cite{Lillo-Mantegna-2000-PRE}. Despite of such remarkable
similarities shared by the ensemble returns and the returns and by
the varieties and the volatilities, there are significant difference
between these ``competing'' quantities, especially the shapes of the
corresponding distributions.

There are a huge number of studies in the literature showing that
emerging stork markets behave differently other than the developed
markets in many aspects. In most developed markets, the daily
returns have well established fat tails, while the distributions of
daily returns are exponential in several emerging markets, e.g., in
China \cite{Wang-Zhang-2005-PA}, Brazil
\cite{Miranda-Riera-2001-PA}, and India
\cite{Matia-Pal-Salunkay-Stanley-2004-EPL}. It is very interesting
to investigate the statistical properties of the ensemble variables
extracted in emerging stock markets, which is the main motivation of
the current work. We shall focus on the Chinese stock markets in
this paper.

The paper is organized as follows. In Sec.~\ref{s1:dataset}, we
explain the data set analyzed and define explicitly the ensemble
return and variety. Section \ref{s1:PDF} presents analysis on the
probability distributions of the daily ensemble returns and
varieties. We discuss in Sec.~\ref{s1:LRD} the temporal correlations
of the two quantities, where we adopt R/S analysis and detrended
fluctuation analysis (DFA) to estimate the Hurst indexes and perform
statistical tests using Lo's modified R/S statistic. The last
section concludes.

\section{China's stock markets}
\label{s1:dataset}

Before the foundation of People's Republic of China in 1949, the
Shanghai Stock Exchange was the third largest worldwide, after the
New York Stock Exchange and the London Stock Exchange and its
evolution over the period from 1919 to 1949 had enormous influence
on other world-class financial markets \cite{Su-2003}. After 1949,
China implemented policies of a socialist planned economy and the
government controlled entirely all investment channels. This proved
to be efficient in the early stage of the economy reconstruction,
especially for the heavy industry. However, planned economic
policies have unavoidably led to inefficient allocation of
resources. In 1981, the central government began to issue treasury
bonds to raise capital to cover its financial deficit, which
reopened the China's securities markets. After that, local
governments and enterprises were permitted to issue bonds. In 1984,
11 state-owned enterprises became share-holding corporations and
started to provide public offering of stocks. The establishment of
secondary markets for securities occurred in 1986 when
over-the-counter markets were set up to trade corporation bonds and
shares. The first market for government-approved securities was
founded in Shanghai on November 26, 1990 and started operating on
December 19 of the same year under the name of the Shanghai Stock
Exchange (SHSE). Shortly after, the Shenzhen Stock Exchange (SZSE)
was established on December 1, 1990 and started its operations on
July 3, 1991. The historical high happened in 2000 when the total
market capitalization reached 4,968 billion yuan (55.5\% of GDP)
with 1,535.4 billion yuan of float market capitalization (17.2\% of
GDP). The size of the Chinese stock market has increased remarkably.

The data set we used in this paper contains daily records of $n=500$
stocks traded in the SHSE and the SZSE in the period from February
1994 to September 2004. The total number of data points exceeds one
million. For each stock price time series, we calculate the daily
log-return as follows
\begin{equation}\label{Eq:return1}
r_i(t)= \ln[P_i(t)/P_i(t-1)]~,
\end{equation}
where $P_i(t)$ is the close price of stock $i$ on day $t$. The
ensemble return $\mu(t)$ is then defined by
\begin{equation}
\mu(t)=\frac{1}{n}\sum_{i = 1}^{n}r_{i}(t)~, \label{Eq:mean}
\end{equation}
while the variety $\sigma(t)$ is defined according to
\begin{equation} \label{Eq:standard}
\sigma^2(t)= \frac{1}{n}\sum_{i = 1}^{n}[r_{i}(t) - \mu(t)]^{2}~.
\end{equation}
The number of active stocks may vary along time $t$. When a stock
$j$ is not traded at time $t$, it is not included in the calculation
of $\mu$ and $\sigma$.

\begin{figure}[htb]
\begin{center}
\includegraphics[width=10cm]{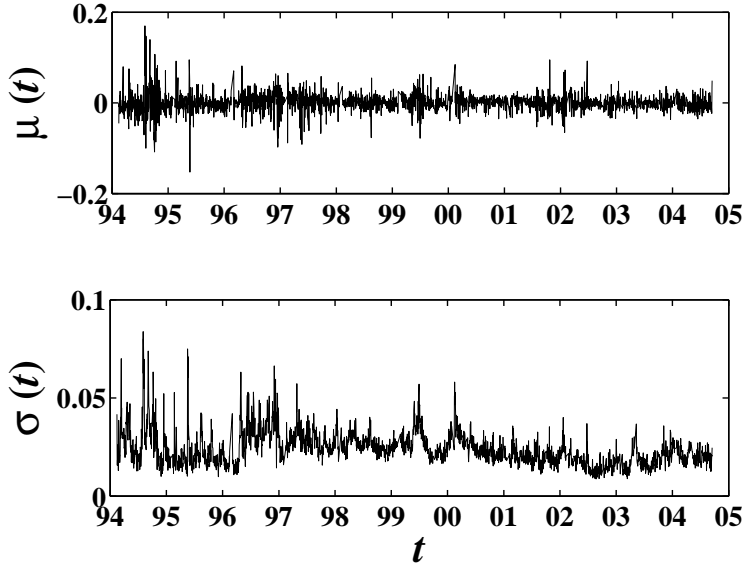}
\end{center}
\caption{Evolution of the daily ensemble return ${\mu}$ (top) and
the daily variety $\sigma$ (bottom) as a function of time.}
\label{Fig:Ensemble}
\end{figure}

Figure~\ref{Fig:Ensemble} illustrates the daily ensemble returns
${\mu}$ and the daily variety ${\sigma}$ in the Chinese stock market
as a function of time $t$ from Feb. 1994 to Sep. 2004. An striking
feature is observed in both quantities that the amplitude of the
envelop decreases along time, which indicates that the Chinese stock
markets are becoming less volatile and more efficient.

\section{Probability distributions of ensemble variables}
\label{s1:PDF}

\begin{figure}[htb]
\begin{center}
\includegraphics[width=10cm]{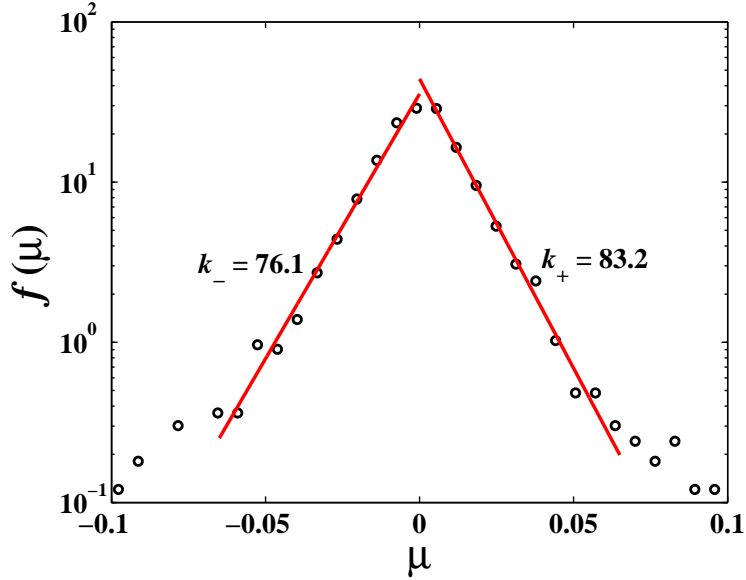}
\end{center}
\caption{Empirical probability density function of ${\mu}$. The
circles represent the real data. The sold lines are the least
squares fits to the exponential (\ref{Eq:PDF:mu:exp}) in the range
$[-0.06,0]$ and $[0,0.06]$ and the slopes are $76.1 \pm 3.3$ and
$83.2 \pm 3.5$, respectively.} \label{Fig:mn_pdf}
\end{figure}

The central parts of ensemble returns of NYSE stocks and Nasdaq
stocks are exponentially distributed and the negative part decays
slower than the positive part
\cite{Lillo-Mantegna-2000-PRE,Lillo-Mantegna-2001-PA}, while the
tails look like outliers in the sense that those ensemble returns
are extremely large and can not be modeled by the same exponential
distribution as the center part \cite{Sornette-2003}. The Chinese
stock markets have the same behavior qualitatively. Figure
\ref{Fig:mn_pdf} shows the empirical probability density function of
${\mu}$. We find that the main part of the density function has the
following form
\begin{equation}\label{Eq:PDF:mu:exp}
    f(\mu) \sim {\exp}[{-k_{\pm}\mu}]~,
\end{equation}
where $k_-=76.1 \pm 3.3$ when $-0.06<\mu<0$ and $k_+=83.2 \pm 3.5$
when $0 < \mu < 0.06$, which shows that the distribution is
asymmetric with the skewness being 0.378. It is interesting to note
that the Chinese stock markets have more large ensemble returns than
the USA markets, which is consistent with the fact that the Chinese
stock markets are extraordinarily volatile.

In order to exploit the tail distribution of the ensemble returns,
we adopt the rank-ordering approach
\cite{Laherrere-Sornette-1998-EPJB,Sornette-Knopoff-Kagan-Vanneste-1996-JGR}.
We first sort the $n$ observations in non-increasing order, that is
$\mu_1 \geq \mu_2 \geq \cdots \geq \mu_R \geq \cdots \geq \mu_n$,
where $R$ is the rank of the observations. Let
$C(\mu)=\int_\mu^\infty f(\mu){\rm{d}}\mu$, then we have
\begin{equation}
nC(\mu_R) = R \label{Eq:nP}
\end{equation}
When the probability density of the ensemble variable $\mu$ scales
as $f(\mu) \sim \mu^{-({1 + \alpha})}$ in the tail, we have
\cite{Laherrere-Sornette-1998-EPJB,Sornette-Knopoff-Kagan-Vanneste-1996-JGR}
\begin{equation}
\mu_R \sim R^{-1/{\alpha}} \label{Eq:xR2}
\end{equation}
for $R\ll N$. A rank-ordering plot of $\ln{\mu_R}$ against $\ln{R}$
thus allows us to check if the tails have power law form.

\begin{figure}[htb]
\begin{center}
\includegraphics[width=10cm]{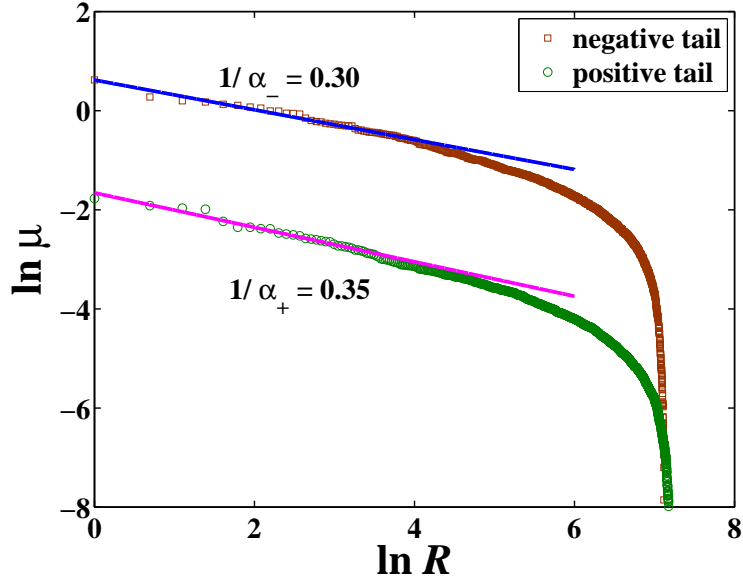}
\end{center}
\caption{Log-log plot of the rank-ordering $\ln{\mu}$ as a function
of $\ln{R}$ for both positive and negative tails(shown in the
legend). The plot of the negative tail is translated vertically for
clarity. The solid lines are the least squares fit to the data at
the interval $0 \leq \ln{R} \leq 4$ for the negative tail and $0
\leq \ln{R} \leq 3.5$ for the positive tail, and the slopes of the
two lines are $0.3 \pm 0.005$ and $0.35 \pm 0.009$, respectively.}
\label{Fig:mn_ro}
\end{figure}

Figure \ref{Fig:mn_ro} shows the rank-ordering plot of both positive
and negative tails, which are approximately power laws. The fitted
tail exponents are $\alpha_- = 3.33 \pm 0.06$ for the negative $\mu$
and $\alpha_+ = 2.86 \pm 0.07$ for positive $\mu$. This is
reminiscent of the inverse cubic law of returns
\cite{Gopikrishnan-Meyer-Amaral-Stanley-1998-EPJB,Gabaix-Gopikrishnan-Plerou-Stanley-2003-PA,Gabaix-Gopikrishnan-Plerou-Stanley-2003-Nature}.

\begin{figure}[htb]
\begin{center}
\includegraphics[width=6.5cm,height=5.5cm]{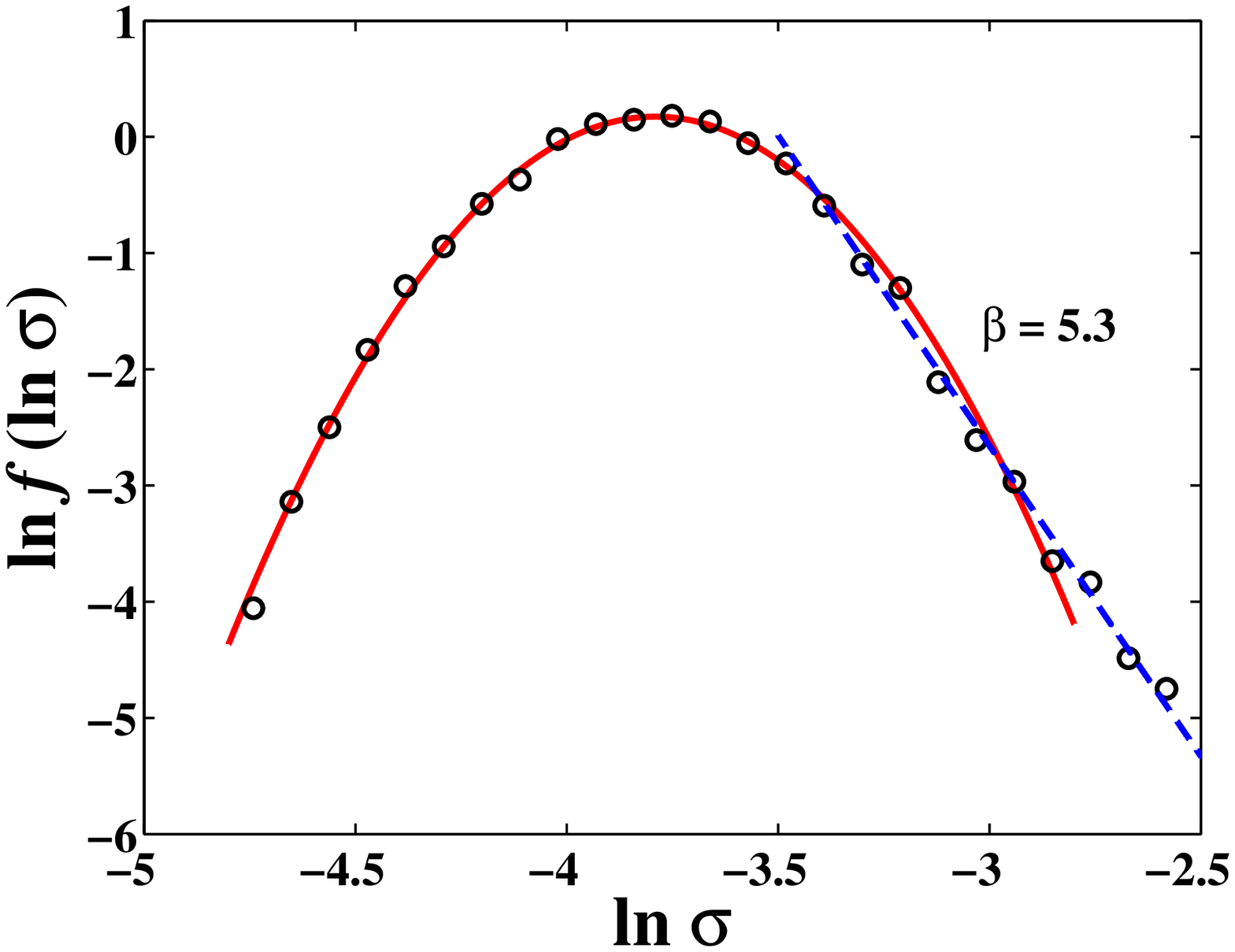}
\includegraphics[width=6.5cm,height=5.5cm]{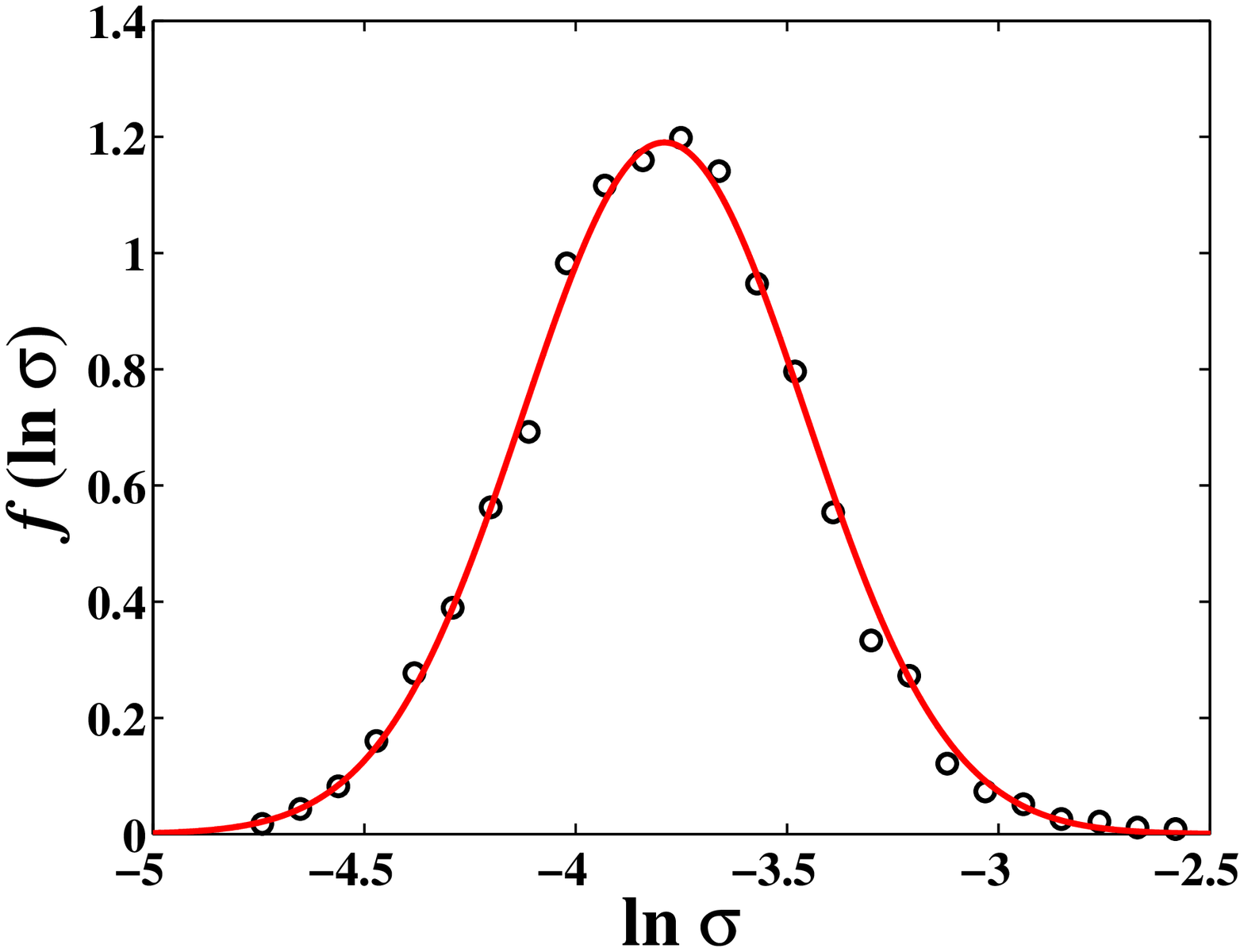}
\end{center}
\caption{Left: Empirical probability density function of the variety
$\sigma$ in double logarithmic coordinates. The continuous line is a
parabolic fit while the dashed line shows a power-law distribution
of larger varieties ($\ln \sigma>-3.5$). The tail exponent is $\beta
= 5.3 \pm 0.2$. Right: Log-normal distribution of ${\sigma}$. }
\label{Fig:pdf:var}
\end{figure}

In Fig.~\ref{Fig:pdf:var} is shown the distribution of varieties of
the Chinese stock markets. It is evident that the main part of the
distribution follows a log-normal form followed by a well
established power law tail:
\begin{equation}\label{Eq:PDF:var}
    f(\sigma) \sim \left\{
    \begin{array}{ll}
      \exp\left[-\frac{(\ln\sigma-\ln\sigma_0)^2}{2{\rm{Var}}(\sigma)}\right], & {\rm{~~~for~not~large~\sigma}} \\
      \sigma^{-\beta}, & {\rm{~~~for~large~\sigma}}
    \end{array}\right.
\end{equation}
where the tail exponent is found to be $\beta = 5.3 \pm 0.2$. Again,
the shape of the variety distribution in the Chinese stock markets
is qualitatively the same as in the USA stock markets
\cite{Lillo-Mantegna-2000-PRE}. Note that the volatilities of most
stock markets have log-normal distributions with power-law tails
\cite{Liu-Gopikrishnan-Cizeau-Meyer-Peng-Stanley-1999-PRE}. However,
the tail distribution of varieties in China's stock markets deviates
from the inverse cubic law.

\section{Long memory in the ensemble variables}
\label{s1:LRD}

\subsection{Detrended fluctuation analysis}

There are a lot of methods developed to extract temporal correlation
in time series, among which the detrended fluctuation analysis (DFA)
is the most popular method due to its easy implementation and robust
estimation even for short time series
\cite{Taqqu-Teverovsky-Willinger-1995-Fractals,Montanari-Taqqu-Teverovsky-1999-MCM,Audit-Bacry-Muzy-Arneodo-2002-IEEEtit}.
DFA was invented originally to study the long-range dependence in
coding and noncoding DNA nucleotides sequence
\cite{Peng-Buldyrev-Havlin-Simons-Stanley-Goldberger-1994-PRE} and
then applied to various fields including finance. In order to
investigate the dependence nature of ensemble variables in China's
stock markets, we first adopt the detrended fluctuation analysis.

The DFA is carried out as follows. Consider a time series $x(t)$,
$t=1,2,\cdots,N$. We first construct the cumulative sum
\begin{equation}
u(t) = \sum_{i = 1}^{t}{x(i)}, ~~t=1,2,\cdots,N~.\label{Eq:dfa_mn_u}
\end{equation}
The time interval is then divided into disjoint subintervals of a
same length $s$ and fit ${u(t)}$ in each subinterval with a
polynomial function, which gives ${u_s(t)}$, representing the trend
in the subintervals. The detrended fluctuation function $F(s)$ is
then calculated
\begin{equation}
F^2(s) = \frac{1}{N}\sum_{i = 1}^{N}[{u(i) - u_s(i)}]^{2}~.
\label{Eq:mn_dfa_F}
\end{equation}
Varying $s$, one is able to determine the scaling relation between
the detrended fluctuation function ${F(s)}$ and time scale $s$. It
is shown that
\begin{equation}
F(s) \sim s^{H}~, \label{Eq:DFA:H}
\end{equation}
where $H$ is the Hurst index
\cite{Taqqu-Teverovsky-Willinger-1995-Fractals,Kantelhardt-Bunde-Rego-Havlin-Bunde-2001-PA},
which is shown to be related to the power spectrum exponent $\eta$
by $\eta=2H-1$
\cite{Talkner-Weber-2000-PRE,Heneghan-McDarby-2000-PRE} and thus to
the autocorrelation exponent $\gamma$ by $\gamma=2-2H$.

\begin{figure}[htb]
\begin{center}
\includegraphics[width=10cm]{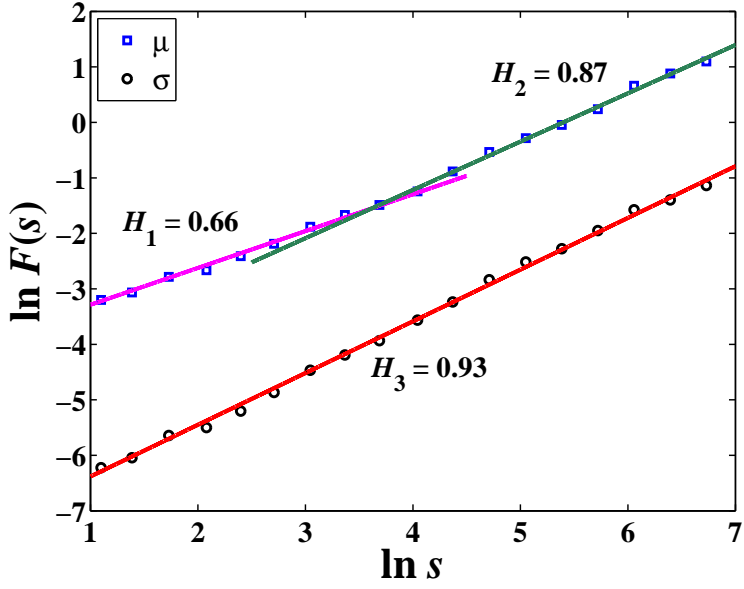}
\end{center}
\caption{Log-log plot is of the detrended fluctuation functions
$F(s)$ of the ensemble daily variables ${\mu}$ and ${\sigma}$ with
respect to the time scale $s$. The squares stand for the results
calculated from real data of ${\mu}$ and the circles represent the
real data of ${\sigma}$. The plot for ${\mu}$ is translated
vertically for clarity. } \label{Fig:dfa}
\end{figure}

Figure \ref{Fig:dfa} plots the detrended fluctuation functions
$F(s)$ of the ensemble daily variables ${\mu}$ and ${\sigma}$ as a
function of time scale $s$. There are two scaling laws in the curve
for ${\mu}$, which are separated at the crossover time lag
$\ln{s_{\times}} = 3.6$. The Hurst indices for both scaling ranges
are $H_1=0.66 \pm 0.02$ and $H_2=0.87 \pm 0.02$, respectively. For
variety ${\sigma}$, a sound power law scaling relation is observed
with a Hurst index $H_3=0.93 \pm 0.01$. This strong correlation
observed here is consistent with that in the USA markets, where the
autocorrelation exponent is reported to be $\gamma=0.230\pm0.006$
\cite{Lillo-Mantegna-2000-PRE}.

\subsection{Rescaled range analysis}

To further investigate the correlation structure in the ensemble
returns and varieties, we adopt the well-known R/S analysis. R/S
analysis was invented by Hurst \cite{Hurst-1951-TASCE} and then
developed by Mandelbrot and Wallis
\cite{Mandelbrot-Wallis-1969b-WRR,Mandelbrot-Wallis-1969d-WRR},
known also as Hurst analysis or rescaled range analysis.

Assume that time series $\{y_i: i=1,2,\cdots,s\}$ is a sub-series
taken from a longer time series $\{x_i:i=1,2,\cdots,N\}$
successively. The cumulative deviation of $\{y_i\}$ is defined by
\begin{equation}
X_{s,i} = \sum_{j=1}^i \left(y_j-\overline{y}\right)~, \label{Eq:x}
\end{equation}
where $\overline{y}$ is the sample average of $\{y_i\}$, and the
range is given by
\begin{equation}
R_{s} = \max_{1\leq i \leq s}X_{s,i} - \min_{1\leq i \leq s}X_{s,i}
\label{Eq:R}
\end{equation}
For a time series with long memory, the range rescaled by the sample
standard deviation
\begin{equation}
S_{s} = \left[{\frac{1}{n}\sum_{i = 1}^{s}{(y_j -
\overline{y})^{2}}}\right]^{1/2} \label{Eq:s}
\end{equation}
scales as a power law with respect to the time scale $s$
\begin{equation}
\frac{R(s)}{S(s)} \sim s^{H}~, \label{Eq:h}
\end{equation}
when $s\to \infty$. There are different algorithms to implement the
R/S analysis, based on the partition of sub-series $\{y_i\}$. We
adopt an algorithm based on random choices of sub-series of size $s$
and averaging over them \cite{Zhou-Liu-Gong-Wang-Yu-2005}.

\begin{figure}[htb]
\begin{center}
\includegraphics[width=10cm]{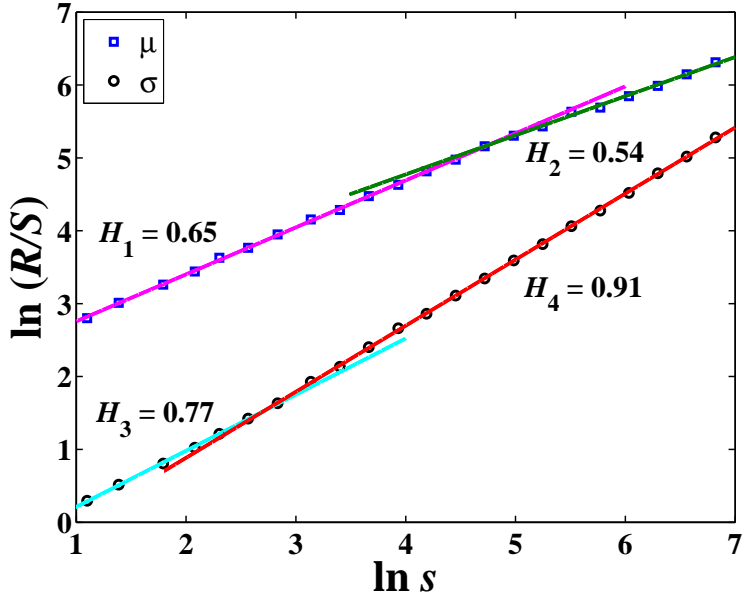}
\end{center}
\caption{R/S analysis of the daily ensemble returns and the daily
varieties. The squares stand for the results calculated from real
data of ${\mu}$ and the circles represent the real data of
${\sigma}$. The two lines are the least squares fit to their results
respectively. The plot for ${\mu}$ is translated vertically for
clarity.} \label{Fig:rs}
\end{figure}

The results of the R/S analysis on the daily ensemble returns and
varieties are presented in Fig.~\ref{Fig:rs}. We observe that both
variables exhibit two scaling ranges. For the ensemble returns, the
crossover occurs at $\ln{s_{\times}} = 4.7$, which should be
compared with the crossover at $\ln{s_{\times}} = 3.6$ in
Fig.~\ref{Fig:dfa}. The Hurst index for small $s$ is
$H_1=0.65\pm0.004$, which is very close to $H_1=0.66\pm0.02$ in
Fig.~\ref{Fig:dfa}. For larger $s$, we have $H_2=0.54 \pm 0.01$,
which is much smaller than $H_2=0.87 \pm 0.02$ in the detrended
fluctuation analysis. This calls for further investigate of possible
long memory in the daily ensemble returns. For the daily, varieties,
the crossover takes place at $\ln{s_{\times}} = 2.8$. The Hurst
index for $s\leqslant{s_{\times}}$ is $H_3=0.77 \pm 0.01$, while for
$s\geqslant{s_{\times}}$ we get $H_4=0.91 \pm 0.003$, which is
consistent with $H_3=0.93\pm0.01$ in the detrended fluctuation
analysis illustrated in Fig.~\ref{Fig:dfa}.

\subsection{Statistical tests of long memory}

The information extracted from the DFA and the R/S analysis
performed on the variety is consistent, where both methods give a
large value of Hurst index. However, the situation is quite
different when the ensemble is concerned. The Hurst indexes for
large time scale $s$ obtained from the two methods are both not far
away from $H=0.5$. Due to the subtlety of the issue of long memory,
we provide further statistical tests for both ensemble variables,
adopting Lo's modified R/S statistic \cite{Lo-1991-Em}.

Consider a stationary time series of size $n$. The modified R/S
statistic is given by \cite{Lo-1991-Em}
\begin{equation}
Q_{n} = R_{n} / \hat{S}_{n}(q)~, \label{Eq:Qn}
\end{equation}
where $R_{n}$ is the range of cumulative deviations defined in
Eq.~(\ref{Eq:R}) and $\hat{S}_{n}(q)$ is defined by
\begin{equation}
\hat{S}_{n}^2(q) = S_{n}^2 + 2\sum_{j = 1}^{q}\omega_{j}(q)\rho_j =
S_{n}^2 + 2\sum_{j = 1}^{q}\left(1 - \frac{j}{q - 1}\right)\rho_j~,
\label{Eq:sigma}
\end{equation}
where $S_n$ is the standard deviation defined in Eq.~(\ref{Eq:s})
and $\rho_{j}=\frac{1}{n}\sum_{i = j + 1}^{n}(y_{i} -
\overline{y})(y_{i-j} - \overline{y})$ is the autocovariance of the
time series. If the time series has short-term memory, the statistic
variable
\begin{equation}
V_n(q) = \frac{1}{\sqrt{n}}Q_n(q) \label{Eq:VV}
\end{equation}
has a finite positive value whose cumulative distribution reads
\begin{equation}
P(V) = 1 +  2\sum_{k = 1}^{\infty}(1 - 4k^2V^2){\rm{e}}^{k^2V^2}~.
\label{Eq:FV}
\end{equation}
The fractiles can be estimated from Eq.~(\ref{Eq:FV}): for a
bilateral test of 5\% significance, we have $V_{0.025}=0.809$ and
$V_{0.975}=1.862$.
When the time series has long-term memory, it is proved that $R_{n}$
trends to the Brownian bridge variable $V_{H}$, while the variable
$S_{n}^{2}/(n\hat{S}_{n}(q))$ tends to $0$ or $\infty$ for large
$q$, that is
\begin{equation}
V_n(q) = \frac{1}{\sqrt{n}}Q_n(q) =
\left\{
\begin{array}{ll}
0, & ~~~~~H\in(0,0.5)\\ \infty, & ~~~~~H\in(0.5, 1)
\end{array} \right.
\label{Eq:Vn}
\end{equation}

These properties allow us to distinguish short memory from long
memory. The null hypothesis and its alternative hypothesis may be
expressed by
\begin{quote}
  $H_0$: The time series under consideration has short memory;\\
  $H_1$: The time series under consideration has long memory.
\end{quote}
The test is performed at the $\alpha$ significant level to accept or
reject the null hypothesis according to whether $V_n(q)$ is
contained in the interval $[V_\alpha,V_{1-\alpha}]$ or not, where
$F(V_\alpha) = \alpha / 2$ and $F(V_{1-\alpha})=1-\alpha/2$. When
$V_n(q)\notin [V_\alpha,V_{1-\alpha}]$, the null hypothesis $H_0$
can be rejected such that the time series has long memory.

We have used $q = 36$, $72$, $108$, and $144$ in the tests. The
tests are performed on the whole time series from 1994/02/14 to
2004/09/15 and its subintervals. The results for ${\mu}$ are
presented in Table \ref{Tab:V_mn}. For the whole time series, the
hypothesis that there is no long memory can not be rejected.
However, the alternative long memory in several subintervals is
significant at the $\alpha=5\%$ level. It is thus possible there
exists long memory in ${\mu}$ in the Chinese stock markets
intermittently, which is not unreasonable due to the inefficiency of
the emerging markets.

\begin{table}[h!]
\begin{center}
\caption{Statistical test of long memory in the daily ensemble
return ${\mu}$ using the modified R/S statistic $V_{n}(q)= Q_n(q) /
{\sqrt{n}}$, which is compared with the classical R/S statistic
$\widetilde{V}_n= \widetilde{Q}_n/\sqrt{n}$, where $\widetilde{Q}_n
= R / S$. } \label{Tab:V_mn}
\begin{tabular}{ccccccc}
\hline
\hline
Time Period & $n$ & $\widetilde{V}_n$ & $V_n(36)$ & $V_n(72)$ & $V_n(108)$ & $V_n(144)$\\
\hline
1994/02/14 - 2004/09/15 & 2568 & 1.81         & 1.66         & 1.70         & 1.70         & 1.72     \\
1994/02/14 - 1999/05/18 & 1284 & $2.11^\star$ & $2.03^\star$ & $2.17^\star$ & $2.19^\star$ & $2.16^\star$ \\
1999/05/19 - 2004/09/15 & 1284 & $1.93^\star$ & 1.55         & 1.47         & 1.46         & 1.50     \\
1994/02/14 - 1996/09/19 & 642  & $2.38^\star$ & $2.35^\star$ & $2.55^\star$ & $2.66^\star$ & $2.88^\star$ \\
1996/09/20 - 1999/05/18 & 642  & $2.00^\star$ & $1.88^\star$ & $2.00^\star$ & $2.09^\star$ & $1.96^\star$ \\
1999/05/19 - 2002/01/14 & 642  & $2.45^\star$ & 1.85         & 1.73         & 1.71         & 1.73     \\
2002/01/15 - 2004/09/15 & 642  & $2.89^\star$ & $2.81^\star$ & $2.77^\star$ & $3.01^\star$ & $3.47^\star$ \\
\hline \hline
\end{tabular}
\end{center}
\end{table}

Table \ref{Tab:V_sd} presents the tests for the daily varieties
${\sigma}$. The long memory hypothesis is significant at the
$\alpha=5\%$ level for all values of $q$ in all subintervals
investigated. For the whole time series, the null hypothesis $H_0$
is rejected for $q=36$ and $q=72$. For larger values of $q$, the
tests show that there is no significant long memory. Since the
definition of the statistic $V_q(n)$ amounts to remove
``autocorrelation'' up to $q$ trading days, the modified R/S test is
biased to over-reject long memory
\cite{Teverovsky-Taqqu-Willinger-1999-JSPI}. Therefore, we argue
that the daily varieties ${\sigma}$ are long-term correlated.

\begin{table}[h!]
\begin{center}
\caption{Statistical test of long memory in the daily ensemble
return using ${\sigma}$ the modified R/S statistic $V_{n}(q)$, which
is compared with the classical R/S statistic $\widetilde{V}_n$.}
\label{Tab:V_sd}
\begin{tabular}{ccccccc}
\hline
\hline
Time Period & $n$ & $\widetilde{V}_n$ & $V_n(36)$ & $V_n(72)$ & $V_n(108)$ & $V_n(144)$\\
\hline
1994/02/14 - 2004/09/15 & 2568 & $11.18^\star$ & $2.67^\star$  & $2.07^\star$ & 1.81     & 1.64     \\
1994/02/14 - 1999/05/18 & 1284 & $19.17^\star$ & $5.43^\star$  & $4.50^\star$ & $4.12^\star$ & $3.85^\star$ \\
1999/05/19 - 2004/09/15 & 1284 & $20.11^\star$ & $4.43^\star$  & $3.48^\star$ & $3.08^\star$ & $2.84^\star$ \\
1994/02/14 - 1996/09/19 & 642  & $19.61^\star$ & $5.70^\star$  & $4.82^\star$ & $4.59^\star$ & $4.49^\star$ \\
1996/09/20 - 1999/05/18 & 642  & $30.15^\star$ & $8.77^\star$  & $7.66^\star$ & $7.27^\star$ & $6.94^\star$ \\
1999/05/19 - 2002/01/14 & 642  & $27.09^\star$ & $6.09^\star$  & $4.99^\star$ & $4.67^\star$ & $4.49^\star$ \\
2002/01/15 - 2004/09/15 & 642  & $42.99^\star$ & $11.52^\star$ & $9.49^\star$ & $8.61^\star$ & $8.14^\star$ \\
\hline
\hline
\end{tabular}
\end{center}
\end{table}

\section{Conclusion}
\label{s1:conclusion}

The ensemble variables ${\mu}$ and ${\sigma}$ are important for
studying the behavior of financial markets as a whole complex
system, instead of individual stocks. In this paper, we have studied
the statistical properties of the daily ensemble returns and daily
varieties of 500 stocks traded in the Shanghai Stock Exchanges and
the Shenzhen Stock Exchanges from 1994/02/14 to 2004/09/15.

The daily ensemble returns ${\mu}$ are found to have exponential
distributions followed by power-law tails. The negative ensemble
returns decay more slowly than the positive part. The negative and
positive tail exponents are $\alpha_-=3.33 \pm 0.06$ and
$\alpha_+=2.86 \pm 0.07$. On the other hand, the daily varieties
${\sigma}$ exhibit a log-normal distribution for not large values
and a power-law form on the tail for large values. The tail exponent
is estimated to be $\beta=5.3\pm0.2$.

There are numerous controversies on the efficiency of the Chinese
stock markets, with slight bias to inefficiency \cite{Su-2003}.
Using detrended fluctuation analysis, R/S analysis and modified R/S
analysis, we have shown that the daily ensemble returns have
long-term memory in several time periods, which is nevertheless
insignificant in the whole time series. Specifically, the long
memory disappears only in the time period from 1999/05/19 to
2002/01/14. This indicates that the Chinese stock markets do not
follow random walks in most time periods. The long-term memory in
the daily varieties is quite strong with a large hurst index
$H=0.91\sim0.93$.

\bigskip
{\textbf{Acknowledgments:}}

This work was supported by the Natural Science Foundation of China
through Grant 70501011.

\bibliography{Bibliography}

\end{document}